\documentclass[useAMS,usenatbib,usegraphicx]{mn2e}

\usepackage{color}

\title[Influence of AGN jets on the magnetized ICM]
  {Influence of AGN jets on the magnetized ICM}

\author[Y. Dubois et al.]
  {Yohan~Dubois,$^1$\thanks{E-mail: yohan.dubois@physics.ox.ac.uk}
  Julien~Devriendt,$^1$ Adrianne~Slyz$^1$ and Joseph~Silk$^1$ \\
  $^1$Astrophysics, University of Oxford, Denys Wilkinson Building, Keble Road, Oxford, OX1 3RH, United Kingdom}
\date{Released 2009 Xxxxx XX}

\pagerange{\pageref{firstpage}--\pageref{lastpage}} \pubyear{2009}

\def\LaTeX{L\kern-.36em\raise.3ex\hbox{a}\kern-.15em
    T\kern-.1667em\lower.7ex\hbox{E}\kern-.125emX}

\begin{document}

  \label{firstpage}

  \maketitle

\begin{abstract}
  Galaxy clusters are the largest structures for which there is
  observational evidence of a magnetised medium. Central cores seem to
  host strong magnetic fields ranging from a few $0.1\, \mu$G up to
  several $10\, \mu$G in cooling flow clusters. Numerous clusters
  harbor central powerful AGN which are thought to
  prevent cooling flows in some clusters. The influence of
  such feedback on the magnetic field remains unclear: does the
  AGN-induced turbulence compensate the loss of magnetic
  amplification within a cool core? And how is this turbulence sustained
  over several Gyr? Using high resolution magneto-hydrodynamical
  simulations of the self-regulation of a radiative cooling cluster, we
  study for the first time the evolution of the magnetic field
  within the central core in the presence of a powerful AGN jet. It appears
  that the jet-induced turbulence strongly amplifies the magnetic amplitude in the core beyond
  the degree to which it would be amplified by pure compression in the gravitational field of the cluster.
  The AGN produces a non-cooling core and increases the
  magnetic field amplitude in good agreement with $\mu$G field
  observations.
\end{abstract}

\begin{keywords}
  galaxies: clusters: general -- galaxies: cooling flows -- galaxies:
  jets -- galaxies: magnetic fields -- methods:
  numerical
\end{keywords}

%%%%%%%%%%%%%%%%%%%%%%%%%%%%%%%%%%%%%%%
% Introduction
%%%%%%%%%%%%%%%%%%%%%%%%%%%%%%%%%%%%%%%
\section{Introduction}

There is increasing evidence for the strong magnetization of the hot
plasma of the intra-cluster medium (ICM, see the review of
\citealt{govoni&feretti04}), the largest scale magnetic field that has been
constrained. But the evolution and the origin of cosmological magnetic
fields remain open questions. Cosmological magnetic fields
could have a primordial origin or could be propagated by galactic winds
emerging from galaxies with strong dynamos. The lack of observational
information on the magnetisation of the inter-galactic medium keeps us
from answering these questions.

Hopefully, numerical simulations may shed light on these
issues. It is starting to become clear that magnetic fields in galaxy
clusters are  compressed essentially by the gravitational collapse of
the gas, and are substantially amplified by the shear motions
developed in this hot phase by the turbulent motions of the
shock-heated surfaces, the galaxy motions and minor/major mergers
\citep{roettigeretal99, dolagetal99, dolagetal02, dolagetal05,
  bruggenetal05, subramanianetal06, asaietal07, dubois&teyssier08cluster}. Mean magnetic
fields inside cluster cores range from a few $0.1\, \mu$G up to
several $10\, \mu$G for the most massive ones, reflecting the 
natural scaling between magnetic field and density.
There is also a difference between magnetic fields in non-cool cores
\citep{kimetal91, clarkeetal01} and cool cores \citep{taylor&perley93,
  vogt&ensslin03, ensslin&vogt06}: cooling flows seem to be a driver
of the magnetic amplification \citep{carilli&taylor02}.

Cosmological numerical simulations are able to reproduce these trends,
adiabatic simulations from \citet{dolagetal05} showing that the largest
clusters are permeated with the strongest magnetic fields. Radiative
simulations from \citet{dubois&teyssier08cluster} have demonstrated that a
cooling flow within a galaxy cluster leads to a dense cool core and
increases the quantity of turbulence and magnetisation in the inner
100 kpc, thereby explaining the fundamental difference between cool cores and
non-cool cores. But the absence of cooling flows in some clusters
is still puzzling as all cores should have already experienced catastrophic collapse \citep{fabian94}.

There are several possible explanations of this puzzle: central
cores could have been heated by thermal conduction from the outer
parts \citep{voigt&fabian04}, or they could have been pre-heated by
stellar outbursts \citep{babuletal02}. But the most popular
explanation is that a central powerful Active Galactic
Nuclei (AGN) is re-heating the core \citep{binney&tabor95, rephaeli&silk95}. 
X-ray surveys show remarkably high energetic outflows in the centres of
galaxy clusters \citep{arnaudetal84, carillietal94, mcnamaraetal01,
  mcnamaraetal05, fabianetal02, birzanetal04, formanetal07}. Such
bursts of energy are associated with supermassive black holes (SMBHs)
accreting hot gas and propelling supersonic jets into the ambient medium
of the ICM plasma \citep{proga03, mckinney06}.

Different attempts have been made to determine the effect of central
feedback on the ICM \citep{bruggen&kaiser02, heinzetal06} and on 
SMBH growth \citep{dimatteoetal05, sijackietal07} in clusters hosting
a powerful AGN. \citet{cattaneo&teyssier07} have pointed out that
accretion onto the central SMBH is a self-regulated process and that
the AGN prevents the formation of a cool core: the jet energy release
in the ICM is sufficient enough to re-heat the ICM.

In this Letter, we  address the question of the evolution of the
magnetic field in the presence of a self-regulated AGN. The exact role of this feedback on cluster core
magnetic fields is still ambiguous. Heating the core might decrease the central density
and thus the magnetic energy, exerting a negative feedback on the
field, but on the other hand the jet could increase the turbulence and
non-negligibly amplify the overall central magnetic field, thereby exerting 
positive feedback.

This problem is studied for the first time in an idealised (but
cosmological) context,  providing new insights on magnetic field
evolution inside galaxy clusters.

%%%%%%%%%%%%%%%%%%%%%%%%%%%%%%%%%%%%%%%
% Numerical considerations
%%%%%%%%%%%%%%%%%%%%%%%%%%%%%%%%%%%%%%%
\section{Numerical considerations}

We are using the Adaptive Mesh Refinement (AMR) code RAMSES
\citep{teyssier02} to solve the full set of ideal
magneto-hydrodynamics (MHD) equations using a constrained transport
method \citep{teyssieretal06, fromangetal06} to preserve the
divergence of the magnetic field. The HLLD solver \citep{miyoshi05}
used in this simulation gives a five wave MHD solution, neglecting the
contribution of the two slow magneto-acoustic modes.

The initial conditions are the same as the ones used in
\citet{cattaneo&teyssier07}, therefore for full details the reader is
invited to refer directly to the latter paper. Let us review the
basics. Dark matter is modelled with a static NFW \citep{navarroetal96} profile
with a concentration parameter $c=5.53$ and a virial mass $M_{\rm
  vir}=1.5\times10^{14}\, \rm M_{\odot}$. The gas density profile is
computed assuming that the gas is in hydrostatic equilibrium with a
baryon fraction $f_{\rm b}=0.1$, and that it follows a polytropic equation
of state with a polytropic index $\gamma_{\rm p}=1.14$. We neglect 
the cluster's rotation, which in any case would be low \citep{bullocketal01}.

We assume that the initial configuration of the magnetic field is the
result of the pure collapse of a gaseous sphere, and that any other
amplification process is negligible. This means that the magnetic
amplitude must follow $B \propto \rho^{2/3}$. To ensure that the
divergence of the field is preserved $\nabla. \vec B=0$, we 
require the magnetic vector $\vec A$
to satisfy $\nabla. (\nabla \times \vec A)=0$ where
\begin{equation}
  {A}(x_i,y_i,z_i)\propto \left ( \begin{array} {c}
      \rho^{2/3}(y_{i},z_{i}) \\
      \rho^{2/3}(x_{i},z_{i}) \\
      0 
    \end{array} \right ) \, ,
\end{equation}
at the centre ($x_i$,$y_i$,$z_i$) of the edge of each cell over the
computational domain. This particular choice of scaling the magnetic
configuration to $\rho^{2/3}$ is justified by the fact that the
magnetic field within the central 200 kpc closely follows the
spherical adiabatic collapse \citep{dubois&teyssier08cluster}. For this study, it
is unimportant if the magnetic field in the outer regions is amplified by shear motions, 
because the jet has a limited range (as
we will see in the next section) and only the central 200 kpc
region is able to radiatively collapse (see
\citealt{dubois&teyssier08cluster}). We must point out that our results are independent of the adopted profile (here $\rho^{2/3}$) as long as field saturation is not reached ($\sim 50 \, \rm \mu$G for a $2.10^7$ K cluster). 
According to this previous paper, we are
fitting the magnetic amplitude in the centre of the cluster to
$10^{-2}\, \mu$G. The cosmological magnetic amplitude in the
intergalactic medium is set to $\simeq 1.3 \times 10^{-5}\, \mu$G at
$\rho=<\rho>$ (i.e. $r\simeq 4$ Mpc), for a fair comparison with
\citet{dolagetal05} and \citet{dubois&teyssier08cluster} that are both
using a $10^{-5}\, \mu$G field in the IGM.

Following the jet implementation in \citet{cattaneo&teyssier07}, the SMBH is a test particle of $M_{\rm SMBH}=3\times 10^9 \, \rm
M_{\odot}$ accreting at a \citet{bondi52} rate $\dot M_{\rm SMBH}=4
\pi (G M_{\rm SMBH})^2 \rho / c_s^3$ (which is a good approximation for X-ray emitting sources, see \citealt{allenetal06}), assuming that the sound speed $c_s
\gg v_{\rm ff}$ where $v_{\rm ff}$ is the infall velocity onto the SMBH. In this model, the SMBH
releases its energy in purely mechanical jet form whenever $\dot M_{\rm SMBH} \ll \dot M_{\rm Edd}$, where $\dot
M_{\rm Edd}$ is the Eddington accretion rate. In this simulation, it is reasonable to follow this
regime since $\dot M_{\rm SMBH} < 0.2\, \rm M_{\odot}/yr$ and $\dot M_{\rm Edd}=65\, \rm M_{\odot}/yr$ (see \citealt{cattaneo&teyssier07}). As we do not resolve the
propagation of the jet, we assume that a certain quantity of the
surrounding gas is entrained with the jet material as it propagates.  We therefore
increase the mass injection rate of the jet $\dot M_{\rm
  J}=\eta \times \dot M_{SMBH}$, by a mass loading factor given by $\eta=100$. 
The jet releases $\epsilon= 10 \%$ of the overall accretion
energy in kinetic form such that the momentum rate of the jet is
$\dot q_{\rm J}=\sqrt{2\epsilon} \dot M_{\rm SMBH} c$ and its rate of energy
deposition is $\dot E_{\rm J}=\epsilon \dot M_{\rm SMBH}c^2$. The
velocity of the jet can therefore be expressed as $u_{\rm J}=\dot q_{\rm J}/ \dot
M_{\rm J}=c\sqrt{2\epsilon}/\eta\simeq1350\, \rm km.s^{-1}$ in
opposite directions. Mass, momentum and energy are spread over a small
cylinder of radius $r_{\rm J}=3.2$ kpc (5 cells) and height of 2$h_{\rm J}$ where $h_{\rm J}=2.5$
kpc (4 cells) multiplied with a kernel window function
\begin{equation}
  \psi={1 \over 2\pi r_{\rm J}^2} \exp \left( -{x^2+z^2\over 2 r_{\rm J}^2 } \right) {y
    \over h_{\rm J}^2} \, ,
\end{equation}
as in \citet{ommaetal04}. Height and radius are arbitrary chosen, but are sufficiently large such that the jet is sampled with several cells. Thus the jet is constrained to flow along the $y$ axis
with only an input $u_y$ velocity component.

The simulation takes into account the self-gravity of the gas within
the dark matter potential. Gas is allowed to cool by radiating
its energy via atomic collisions \citep{sutherland&dopita93}
assuming 75 \% Hydrogen and 25 \% Helium fractions. The hot plasma gas has a $Z=Z_{\odot}/3$ metallicity constant through time and space, 
allowing the gas to cool more efficiently down to $ 10^4$ K.

Our computational domain is $L_{\rm box}=$648 kpc long with a $64^3$
coarse grid resolution ($\ell_{\rm min}=6$). 
We refine the grid according to a geometric strategy (no
time evolution): the $\ell=7$ grid is embedded within a $L_{\rm
  box}/8$ sphere radius centred on the SMBH, $\ell=8$ and $\ell=9$
within a $L_{\rm box}/16$ sphere radius, and $\ell=10$ within a
$L_{\rm box}/32$ sphere radius.
In this way the
resolution within the centre reaches an equivalent $1024^3$ grid
($\ell_{\rm max}=10$) allowing for a minimum cell size $\Delta x
=0.64$ kpc. 

%%%%%%%%%%%%%%%%%%%%%%%%%%%%%%%%%%%%%%%
% Results
%%%%%%%%%%%%%%%%%%%%%%%%%%%%%%%%%%%%%%%
\section{Results}

%%%%%%%%%%%%%%%%%%%%%%%%%%%%%%%%%%%%%%%
% Cooling flow clusters
%%%%%%%%%%%%%%%%%%%%%%%%%%%%%%%%%%%%%%%
\subsection{Cooling flow clusters}

\begin{figure}
  \centering{\resizebox*{!}{7.5cm}{\includegraphics{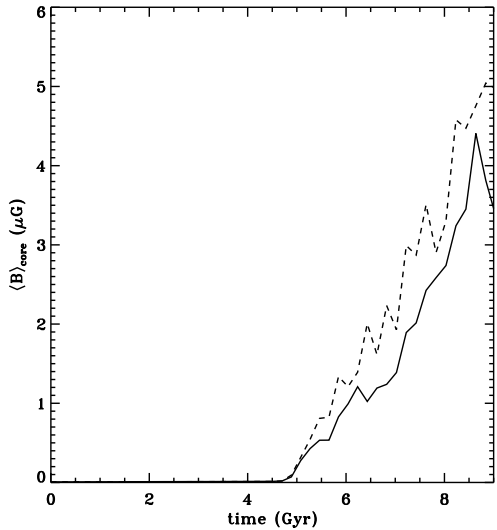}}}
  \caption{Average magnetic amplitude (solid line) in the central core
    $r< 50$ kpc as a function of time for the no-AGN run. A pure
    $\rho^{2/3}$ scaling law is overplotted (dashed line).}
  \label{EBcorevst_nojet}
\end{figure}

We tackle first the case of no AGN in the galaxy cluster.
Since there is nothing to
prevent the core from collapsing, a strong cooling flow develops and the
gas density reaches very high values. We compute the
density--weighted average magnetic field as
\begin{equation}
  <B>_{\rm core}=\left( { \sum_{<r_{\rm core}} B^2 \rho dV \over \sum_{<r_{\rm core}} \rho dV} \right)^{1/2}\, ,
\end{equation}
for each cell lying within the core radius taken to be $r_{\rm core}=50$ kpc. The
definition of the core radius is quite ambiguous. However what is important
here is to properly measure the excess of magnetic amplification relative to
the case of pure collapse. As we weight the average magnetic
amplitude by the density, denser regions contribute more to the
results. Thus taking a larger core radius will not lead to a different
behaviour, but care should be taken that 
the core radius is not so small that
all the effects of the turbulent amplification (in the jet
case) are lost. In the AGN case, we also checked that the magnetic field in the
core is independent of the unresolved region of the jet by excluding
this small region ($<4$ kpc) from the calculation of the density--weighted average magnetic field.

As shown in fig.~\ref{EBcorevst_nojet}, the magnetic field 
rises up to a few $\mu$G in the no-AGN case, because of the
catastrophic collapse occurring at 4 Gyr. As there is no turbulence in
this run, the magnetic field rather nicely follows the evolution expected for pure compression
($ B \propto \rho^{2/3}$ where $\rho$ is the average density). 
The discrepancy between the measured magnetic amplitude and that
predicted by pure compression is due to spurious (numerical)
magnetic reconnection in the centre of the core where the flow
converges.

%%%%%%%%%%%%%%%%%%%%%%%%%%%%%%%%%%%%%%%
% Non-cooling flow clusters
%%%%%%%%%%%%%%%%%%%%%%%%%%%%%%%%%%%%%%%
\subsection{AGN feedback: non-cooling flow clusters}

\begin{figure}
  \centering{\resizebox*{!}{7.5cm}{\includegraphics{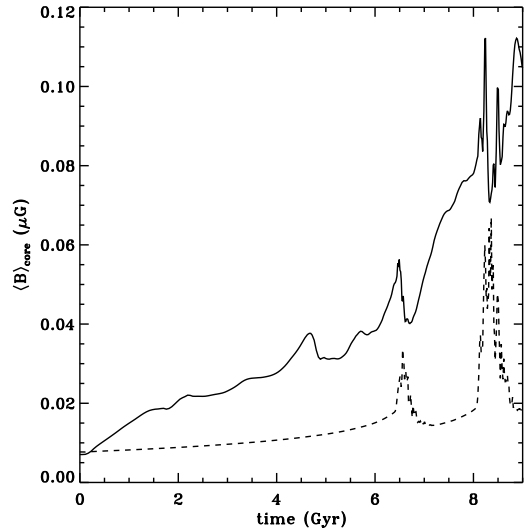}}}
  \caption{Same as fig.~\ref{EBcorevst_nojet} for the AGN run.}
  \label{EBcorevst}
\end{figure}

\begin{figure*}
  \centering{\resizebox*{!}{7cm}{\includegraphics{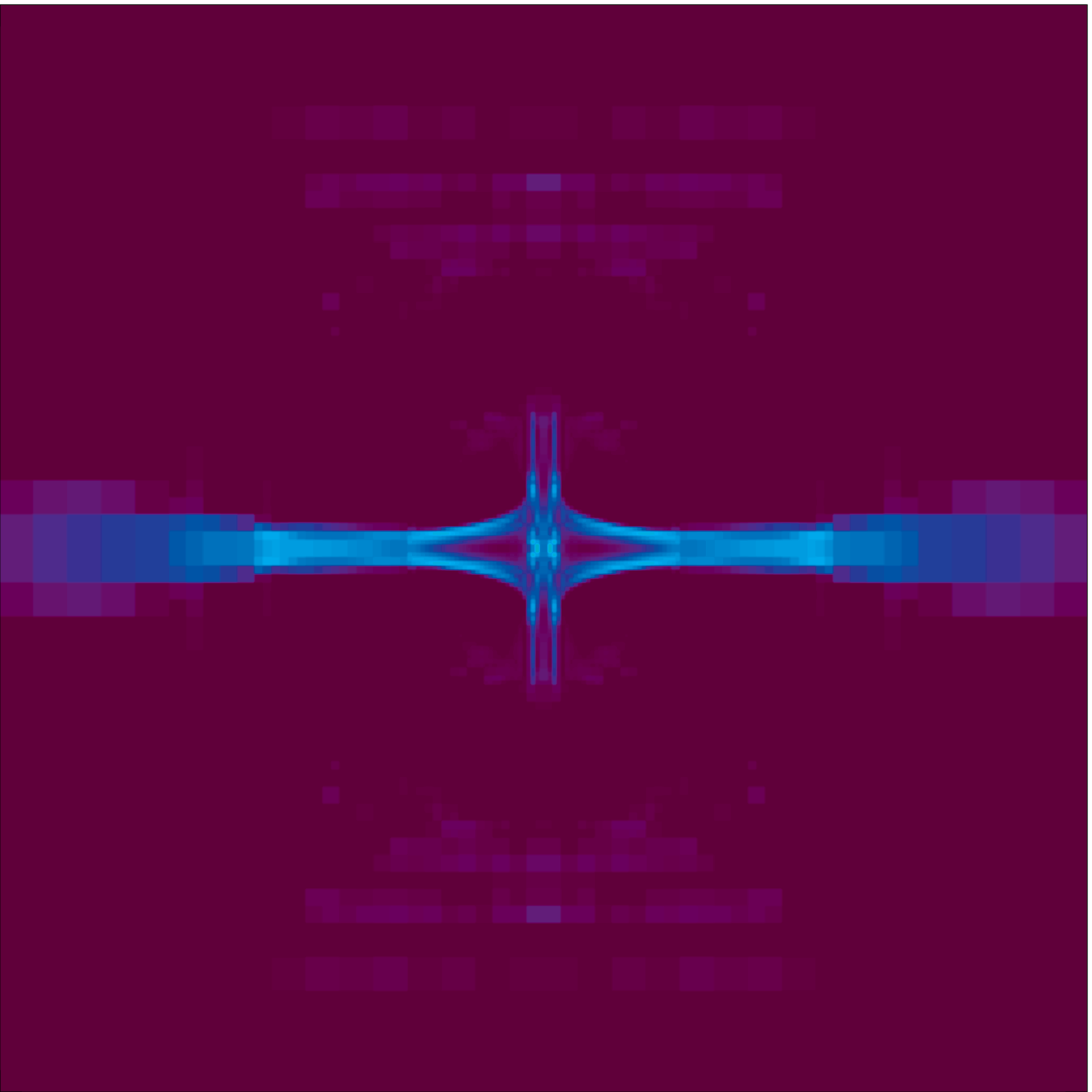}}}
  \centering{\resizebox*{!}{7cm}{\includegraphics{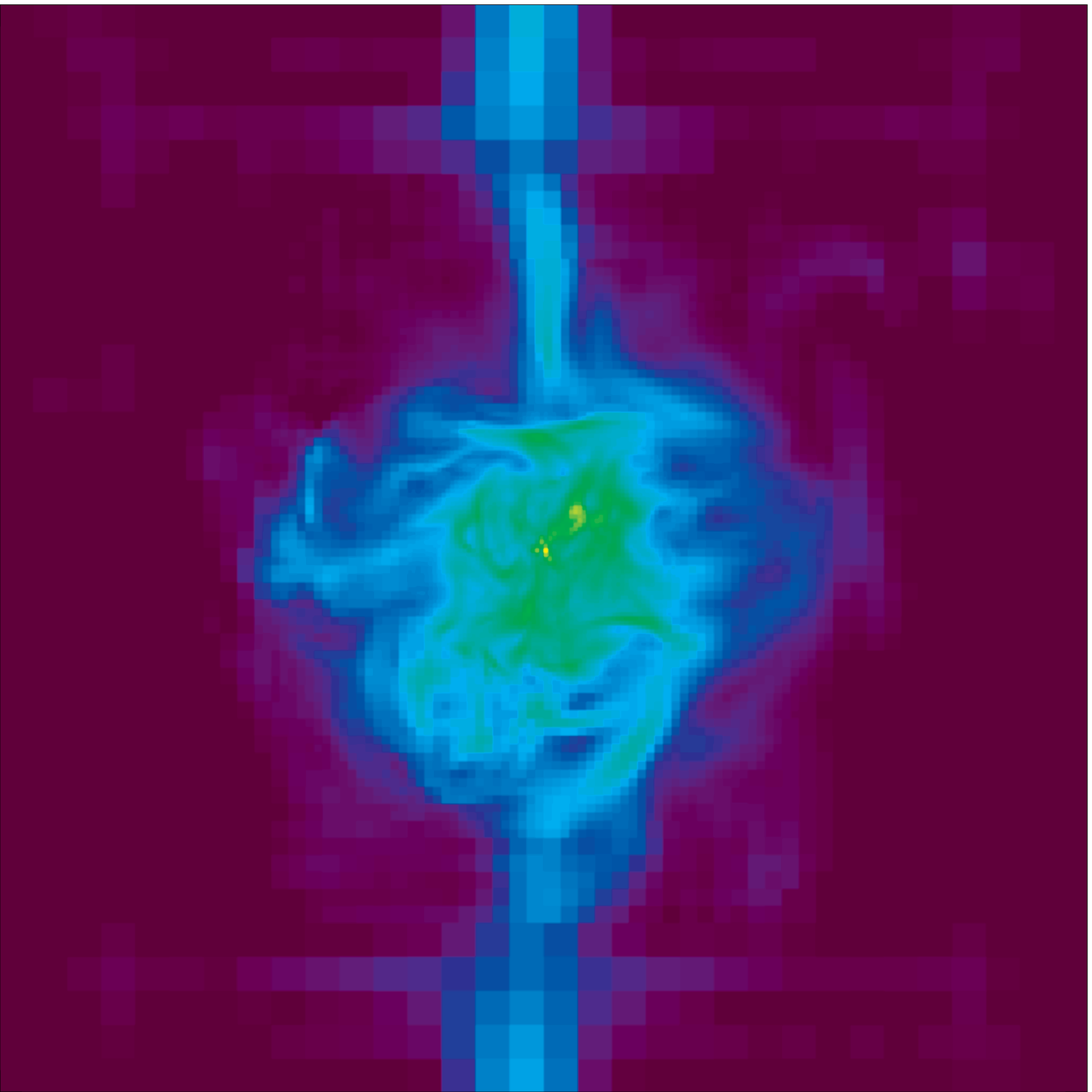}}}
  \centering{\resizebox*{!}{7cm}{\includegraphics{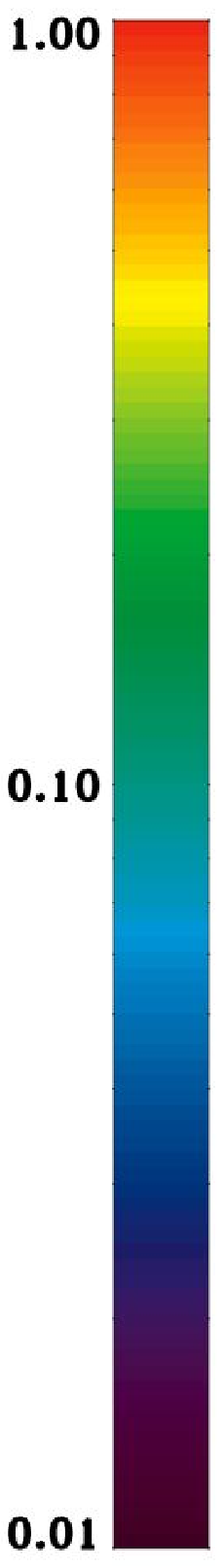}}}
  \caption{Projection along the $x$ axis of the density-weighted
    average of the magnetic amplitude in $\mu$G units after (a) $t=2$
    Gyr (left panel) and (b) $t=8$ Gyr (right panel) for the AGN run. The image size
    is $L_{\rm box}/4=162$ kpc. In the left panel: magnetic field is stretched and amplified along the jet axis. In the right panel: catastrophic collapse has occurred, the jet propagation is no longer straightforward and brings turbulence into the plasma. Thus the magnetic field is filling a small sphere and some magnetic filaments are created due to shear motion of the gas. The vertical magnetic feature is due to the initial configuration of the field (along the $z$ axis). }
  \label{Bmod_proj}
\end{figure*}

If we make the same measurement for the AGN run, the results are fundamentally different
(fig. ~\ref{EBcorevst}): the magnetic amplitude in the core reaches
only $0.1\, \mu$G, but it is strongly amplified relative to the pure
compression regime. Initially the magnetic field is a dipole-like
structure aligned with the $z$ axis, and the jet (aligned with the $y$ axis) is perpendicular
to the field. The jet stretches and
amplifies the $y$ component of the field in a few $100$ Myr up to
$0.1\, \mu$G, as one can see in fig.~\ref{Bmod_proj}(a). The jet
propagates far from the centre of the cluster ($\sim 100$ kpc) until
the ram-pressure of the ICM has sufficiently dissipated its mechanical
energy. The cluster core and its magnetic evolution remains in this
quasi-stationnary state. As there is no initial turbulence in the cluster, the jet
remains nearly unperturbed.

At $\sim 6.5$ Gyr, the cooling catastrophe occurs: the very centre of
the cluster core collapses in a free-fall time. In this case, the jet is 
unable to prevent this critical collapse, but as the density
rises in the centre, the accretion rate onto the SMBH grows and as a consequence
the energy released by the jet increases up to the point that the jet
begins to halt the accretion onto the SMBH.
In this way, the cluster core evolution is self-regulated by its AGN. 
This self-regulated situation is highly unstable, and
alters the propagation direction of the jet, leading to turbulent motions within the cluster core.
Shear motions twist magnetic field lines (fig.~\ref{Bmod_proj}.(b)), so that in the core the
magnetic field is no longer aligned with the initial
field. The relatively strong magnetic amplitude on the $z$ axis, at
$t=8$ Gyr and $z> 100$ kpc, as opposed to that along the $y$ and $x$ axes 
is the result of the preferential orientation of the initial
field (the same situation occurs if there is no AGN feedback).

We remark that the self-regulation process of the AGN is crucial
for the magnetic field evolution. Each time the cluster core suffers a
rapid collapse due to cooling (at $t\simeq6.5$ Gyr and $t\simeq8.2$ Gyr), the
overall magnetic amplitude in the core rapidly grows by
compression. As the SMBH is fed at increasing rates, the AGN is strenghtened, leading to
a violent outburst in the core which inhibits the cooling flow, powers the turbulence, 
and hence the magnetic amplification within the cluster core.

\begin{figure}
  \centering{\resizebox*{!}{7.5cm}{\includegraphics{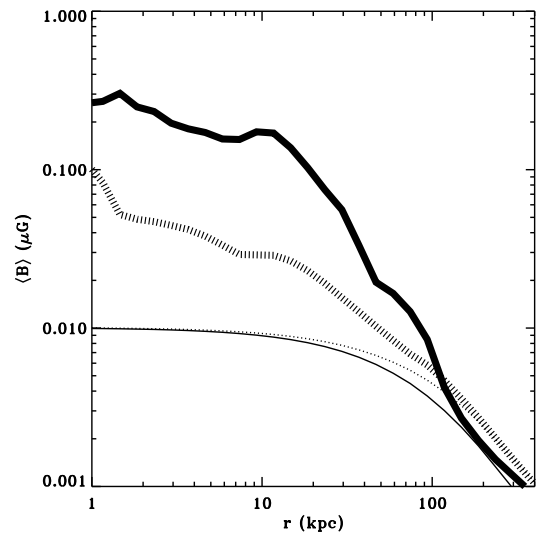}}}
  \caption{Average magnetic amplitude (solid lines) and $\rho^{2/3}$
    law (dotted lines) at $t=0$ Gyr (thin lines) and $t=8$ Gyr (thick
    lines) for the AGN run. One can see that the initial conditions for the magnetic field are fitted on the $\rho^{2/3}$ profile.}
  \label{EBvsr}
\end{figure}

All this amplification is mainly concentrated within the centre of the
cluster as shown by fig.~\ref{EBvsr}. We can see that the magnetic
field at $t=8$ Gyr is strongly amplified (by a factor $\sim4$--$5$) up
to a distance of $\sim 100$ kpc from the centre. As a source of turbulence, the jet
can propagate shear motions only up to that distance, meaning that all
the jet-driven turbulence is concentrated within the core.  The typical 3D velocity dispersion in the core at $t=8$ Gyr is $\sigma\simeq 100-170\, \rm km/s$ in fair agreement with \cite{subramanianetal06}. Finally the AGN is sufficient enough to prevent the creation of a dense and cool
core.

%%%%%%%%%%%%%%%%%%%%%%%%%%%%%%%%%%%%%%%
% Discussion
%%%%%%%%%%%%%%%%%%%%%%%%%%%%%%%%%%%%%%%
\section{Discussion}

These numerical results of the magnetic evolution of a galaxy cluster
with or without an AGN jet are able to explain the discrepancy between strong
magnetic fields in cool cores (no AGN heating) and low magnetic fields in non-cool
cores (AGN heating). Without any AGN, a strong cooling flow appears and the magnetic
amplitude in the cool core ($< 50$ kpc) reaches $\simeq 4\, \mu$G by
gravitational compression alone. The presence of a powerful
AGN jet, on the other hand, leads to softer and hotter cores within which the magnetic amplitude
is $\simeq 0.1 \, \mu$G but turbulence, rather than gravitational compression,  drives the amplification
of the field. 

This work is in good agreement with previous numerical studies. In
particular, \citet{dubois&teyssier08cluster} have shown that adiabatic
cores (no AGN feedback and no cooling) are unable to sufficiently
amplify the magnetic field within the central core ($10^{-2}\, \mu$G),
but cooling flows are able to match observations (several $0.1\,
\mu$G). Here we have shown that AGN are able to stop a central cooling flow
and to prevent the formation of a dense core, and at the same time the
jet-driven turbulence can amplify inner magnetic fields up to a few $\mu$G
in accordance with observations of non-cool cores. 
The scenario of a thermally conducting plasma in the absence of AGN 
feedback seems to be weakened. Thermal conduction can 
re-heat the core, but it does not bring in sufficient 
turbulence  \citep{kim&narayan03}. As we need 
strong turbulent amplification of the initial magnetic field
for non-cool core clusters, it is difficult to explain how 
$\mu$G fields can be reached in scenarios that only rely on thermal conduction.

Of course we are far from reaching the full magneto-turbulent cascade
with the kiloparsec resolution of cosmological simulations. But such
simulations give access to the full evolution of a galaxy
cluster over a fraction of the Hubble time, and allow one to explain the
large-scale evolution of these magnetised structures. We must also
point out that the only source of turbulence in our simulation is the jet
itself, meaning we neglect all the contributions coming from
the outer parts of the cluster. Shock-heated turbulence could be
important, especially in destabilising the jet and forcing it to stay in
the core. We also did not consider any magnetization of the jet that could either
collimate it or magnetize the entire cluster core. Indeed, even though the 
amount of magnetic energy injected into the ICM by AGN jets is yet unknown,
 \cite{xuetal09} have shown that a temporary deposit of magnetic energy 
within the jet, coupled to an efficient turbulent dynamo is able to somewhat 
enrich the hot plasma. We simply claim in this letter that within the interior
of clusters, self-regulated AGN jets are able to sustain turbulence in the long term
and to therefore efficiently amplify/transport magnetic fields.

This Letter is the first attempt to self-consistently consider the
evolution of AGN feedback powering turbulence in a magnetised
cluster. Our work suggests that it is of particular importance to
include AGN feedback if one wants to reproduce observations and explain
the origin and the evolution of the magnetic field on very large
scales of the Universe. In a forthcoming paper, we will address
this problem in a fully non-idealised cosmological context.

\section*{Acknowledgment}
The authors wish to thank Andrea Cattaneo and Romain Teyssier for
giving us access to a public version of their initial conditions. We want
to thank Kumiko Kotera and Stas Shabala for useful discussions. Y. D. is supported by an STFC Postdoctoral Fellowship.
The simulations presented here were run on the
JADE cluster at the Centre d'Informatique National de l'Enseignement
Sup\'erieur in Montpellier.

\bibliographystyle{mn2e} \bibliography{author}

\label{lastpage}

\end{document}